\documentclass{llncs}

\usepackage{etex}
\usepackage[english]{babel} 
\usepackage[latin1]{inputenc} 
\usepackage{latexsym}
\usepackage{epsfig} 
\usepackage{listings}
\usepackage{url,xspace} 
\usepackage[draft]{fixme}
\usepackage{color} 
\usepackage{multirow}
\usepackage{latexsym,amssymb}
\usepackage{fancybox,epic}
\usepackage[noend]{algpseudocode}
\usepackage{amsmath}
\usepackage{graphics}
\usepackage{graphicx}
\usepackage{todonotes}
\usepackage{hyperref}
\usepackage{cite}

\usepackage{graphicx}
\usepackage{xcolor}
\definecolor[named]{ACMBlue}{cmyk}{1,0.1,0,0.1}
\definecolor[named]{ACMYellow}{cmyk}{0,0.16,1,0}
\definecolor[named]{ACMOrange}{cmyk}{0,0.42,1,0.01}
\definecolor[named]{ACMRed}{cmyk}{0,0.90,0.86,0}
\definecolor[named]{ACMLightBlue}{cmyk}{0.49,0.01,0,0}
\definecolor[named]{ACMGreen}{cmyk}{0.20,0,1,0.19}
\definecolor[named]{ACMPurple}{cmyk}{0.55,1,0,0.15}
\definecolor[named]{ACMDarkBlue}{cmyk}{1,0.58,0,0.21}
\hypersetup{colorlinks,
  linkcolor=ACMDarkBlue,
  citecolor=ACMPurple,
  urlcolor=ACMBlue,
  filecolor=ACMDarkBlue}

\newcommand{\point}[1]{\par\smallskip\noindent{\textbf{#1}}.}

\usepackage{epsfig}

\usepackage{booktabs}
 \usepackage[all]{xy}

\usepackage{amssymb}
\usepackage{latexsym}
\usepackage{fancybox,epic}
\usepackage{algorithm}
\usepackage{amsmath}
\usepackage{colortbl}

\usepackage{pgf}
\usepackage{tikz}
\usepackage{amsmath}

\let\llncssubparagraph\subparagraph
\let\subparagraph\paragraph
\usepackage{titlesec}
\let\subparagraph\llncssubparagraph

\titlespacing*{\section}{0pt}{*1}{*1}
\titlespacing*{\subsection}{0pt}{*0.7}{*0.5}
\titlespacing*{\paragraph}{0pt}{*0.5}{*0.5}

\usepackage{listings}

\definecolor{deepblue}{rgb}{0,0,0.5}
\definecolor{deepred}{rgb}{0.6,0,0}
\definecolor{deepgreen}{rgb}{0,0.5,0}
\definecolor{halfgray}{gray}{0.55}
\definecolor{ipythonframe}{RGB}{207, 207, 207}

\definecolor{ckeyword}{HTML}{7F0055}
\definecolor{ccomment}{HTML}{3F7F5F}
\definecolor{cnumber}{HTML}{2A0099}
\definecolor{pblue}{rgb}{0.13,0.13,1}
\definecolor{pgreen}{rgb}{0,0.5,0}
\definecolor{pred}{rgb}{0.9,0,0}
\definecolor{pgrey}{rgb}{0.46,0.45,0.48}

\definecolor[named]{ACMBlue}{cmyk}{1,0.1,0,0.1}
\definecolor[named]{ACMYellow}{cmyk}{0,0.16,1,0}
\definecolor[named]{ACMOrange}{cmyk}{0,0.42,1,0.01}
\definecolor[named]{ACMRed}{cmyk}{0,0.90,0.86,0}
\definecolor[named]{ACMLightBlue}{cmyk}{0.49,0.01,0,0}
\definecolor[named]{ACMGreen}{cmyk}{0.20,0,1,0.19}
\definecolor[named]{ACMPurple}{cmyk}{0.55,1,0,0.15}
\definecolor[named]{ACMDarkBlue}{cmyk}{1,0.58,0,0.21}

\lstdefinelanguage{Solidity} {
  keywords={typeof, modifier, function, public, returns, external,
  contract, new, true, false, private, catch, function, return, null, throw, catch, switch, var, if, in, while, do, else, case, break},
  ndkeywords={bool, address, mapping, uint, bytes32, string},
  identifierstyle=\color{black},
  sensitive=false,
  comment=[l]{//},
  morecomment=[s]{/*}{*/},
  commentstyle=\color{ccomment}\ttfamily,
  string=[b]",
  showstringspaces=false,
  morestring=[b]',
  showspaces=false,
  showtabs=false,
  breaklines=true,
  morekeywords={function, contract, returns, return},
  breakatwhitespace=true,
  lineskip=-0.6pt,
  basewidth={0.54em, 0.4em},
  basicstyle=\footnotesize\ttfamily,
  keywordstyle={\color{ckeyword}\scriptsize\bfseries},
  ndkeywordstyle={\color{pblue}\scriptsize\bfseries},
  commentstyle={\color{ccomment}\itshape},
  stringstyle={\color{pgreen}},
  numberstyle={\scriptsize\color{cnumber}\ttfamily},
  moredelim=[il][\textcolor{pgrey}]{$$},
  moredelim=[is][\textcolor{pgrey}]{\%\%}{\%\%},
}

\newcommand{\scode}[1]{\lstinline[language=Solidity,basicstyle=\small\ttfamily]{#1}}
\newcommand{\code}[1]{\scode{#1}}

\definecolor{shadecolor}{gray}{1.00}
\definecolor{ddarkgray}{gray}{0.75}
\definecolor{darkgray}{gray}{0.30}
\definecolor{light-gray}{gray}{0.87}

\newcommand{\eg}{\emph{e.g.}\xspace}

\newcommand{\tname}[1]{\textsc{#1}\xspace}

\newcommand{\toolname}{\tname{EthIR}} 

\newcommand{\oyente}{\tname{Oyente}}

\newcommand{\translate}{\tau}

\newcommand{\lst}[1]{\lstinline!#1!}

\newcommand{\Get}[1]{\ensuremath{\mbox{\bf \lstinline!get!}}\xspace}

\newcommand{\returnval}[1]{\mbox{\lstinline!return #1!}\xspace}

\newcommand{\extend}[1]{S}

\newcommand{\secbeg}{\vspace*{-0.15cm}}

\newcommand{\lot}[1]{\xoverline{l_{#1}}}
\newcommand{\sot}[1]{\xoverline{s_{#1}}}
\newcommand{\got}[1]{\xoverline{g_{#1}}}
\newcommand{\blot}[0]{\xoverline{bc}}
\newcommand{\inot}[1]{\xoverline{i_{#1}}}
\newcommand{\svar}[1]{s_{#1}}

\newcommand{\lvar}[1]{l_{#1}}

\makeatletter
\newsavebox\myboxA
\newsavebox\myboxB
\newlength\mylenA

\newcommand*\xoverline[2][0.75]{%
    \sbox{\myboxA}{$\m@th#2$}%
    \setbox\myboxB\null
    \ht\myboxB=\ht\myboxA%
    \dp\myboxB=\dp\myboxA%
    \wd\myboxB=#1\wd\myboxA
    \sbox\myboxB{$\m@th\overline{\copy\myboxB}$}
    \setlength\mylenA{\the\wd\myboxA}
    \addtolength\mylenA{-\the\wd\myboxB}%
    \ifdim\wd\myboxB<\wd\myboxA%
       \rlap{\hskip 0.7\mylenA\usebox\myboxB}{\usebox\myboxA}%
    \else
        \hskip -0.5\mylenA\rlap{\usebox\myboxA}{\hskip 0.7\mylenA\usebox\myboxB}%
    \fi}
\makeatother

\lstdefinestyle{numbers}
{numbers=left, numberstyle=\tiny}

\title{{\sc\toolname}: A Framework for High-Level Analysis\\ of Ethereum
  Bytecode
\vspace{-5pt}
}

 \author{Elvira Albert \and   Pablo Gordillo  \and
   Benjamin Livshits  \and\\  Albert Rubio \and Ilya Sergey}

\institute{
\vspace{-10pt}
}

\begin{document}
\maketitle


\begin{abstract}
  Analyzing Ethereum bytecode, rather than the source code from which
  it was generated, is a necessity when: (1) the source code is not
  available (e.g., the blockchain only stores the bytecode), (2) the
  information to be gathered in the analysis is only visible at the
  level of bytecode (e.g., gas consumption is specified at the level
  of EVM instructions), (3) the analysis results may be affected by
  optimizations performed by the compiler (thus the analysis should be
  done ideally \emph{after} compilation). This paper presents
  \toolname, a framework for analyzing Ethereum bytecode, which relies
  on (an extension of) \oyente, a tool that generates CFGs; \toolname
  produces from the CFGs, a \emph{rule-based representation} (RBR) of
  the bytecode that enables the application of (existing)
  high-level analyses to infer properties of EVM code.
\end{abstract}



\section{Introduction}
\label{sec:intro}

Means of creating distributed consensus have given rise to a family of
distributed protocols for building a replicated transaction log (a
\emph{blockchain}). These technological advances enabled the creation
of decentralised cryptocurrencies, such as Bitcoin~\cite{Nakamoto:08}.
Ethereum~\cite{Wood:Ethereum}, one of Bitcoin's most prominent
successors, adds Turing-complete stateful computation associated with
funds-exchanging transactions---so-called \emph{smart contracts}---to
replicated distributed storage.

Smart contracts are small programs stored in a blockchain
that can be invoked by transactions initiated by parties
involved in the protocol, executing some business logic as automatic
and trustworthy mediators. Typical applications of smart contracts
involve implementations of multi-party accounting, voting and
arbitration mechanisms, auctions, as well as puzzle-solving games with
reward distribution.
%
%
To preserve the global consistency of the blockchain, every transaction involving an
interaction with a smart contract is replicated across the system.
%
%
In Ethereum, replicated execution is implemented by means of a
uniform execution back-end---Ethereum Virtual Machine
(EVM)~\cite{Wood:Ethereum}---a stack-based operational formalism,
enriched with a number of primitives, allowing contracts to call each
other, refer to the global blockchain state, initiate
sub-transactions, and even create new contract instances
dynamically. That is, EVM provides a convenient \emph{compilation
  target} for multiple high-level programming languages for
implementing Ethereum-based smart contracts.
In contrast with prior low-level languages for smart
contract scripting, EVM features mutable
persistent state that can be modified, during a contract's lifetime,
by parties interacting with it. 
Finally, in order to tackle the issue of possible denial-of-service
attacks, EVM comes with a notion of \emph{gas}---a cost semantics of
virtual machine instructions.
%

All these features make EVM a very powerful execution formalism,
simultaneously making it quite difficult to formally analyse its
bytecode for possible inefficiencies and vulnerabilities---a challenge
exacerbated by the mission-critical nature of smart contracts, which,
after having been deployed, cannot be amended or taken off the
blockchain.

\point{Contributions} 
In this work, we take a step further towards
\emph{sound} and \emph{automated} reasoning about high-level
properties of Ethereum smart contracts. 
\begin{itemize}
\item 
We do so by providing \toolname, an open-source tool for precise
decompilation of EVM bytecode into a high-level representation in a
rule-based form; \toolname is available via GitHub: \url{https://github.com/costa-group/ethIR}.  
\item 
Our representation reconstructs high-level control and data-flow for EVM bytecode from
the low-level encoding provided in the CFGs generated by \oyente. It
enables application of
state-of-the-art analysis tools developed for high-level languages to
infer properties of bytecode. 
\item 
We showcase this application by 
conducting an automated resource analysis of existing contracts from
the blockchain inferring their loop bounds.
\end{itemize}









 


\section{From EVM to a Rule-based Representation}
\label{sec:decompilation}

The purpose of decompilation --as for other bytecode languages (see,
e.g., the Soot analysis and optimization framework\cite{soot})-- is to
make explicit in a higher-level representation the \emph{control flow}
of the program (by means of rules which indicate the continuation of
the execution) and the \emph{data flow} (by means of explicit
variables, which represent the data stored in the stack, in contract
fields, in local variables, and in the blockchain), so that an
analysis or transformation tool can have this control flow information
directly available.

\subsection{Extension of Oyente to Generate the CFG}\label{sec:oyente}


Given some EVM code, the \oyente tool generates a set of blocks that
store the information needed to represent the CFG of such EVM
code. 
However, when the jump address of a block is not unique (depends on
the flow of the program), the blocks generated by \oyente sometimes
only store the last value of the jump address (this is because it is
enough for the kind of symbolic execution they perform).  We have
modified the structure of \oyente blocks in order to include all
possible jump addresses, so that the whole CFG is reconstructed. As an
example, Fig.~\ref{fig:cfg} shows the Solidity source code
for a fragment of a contract (left), and the CFG generated from
it (right).  Observe that in the CFGs generated by our extension of \oyente,
the instructions SSTORE or SLOAD are annotated with an identifier of
the contract field they operate on (for instance, a SSTORE operation
that stores a value on the contract field 0 is replaced by SSTORE
0). Similarly, the EVM instructions MSTORE and MLOAD instructions are
annotated with the memory address they operate on (such addresses will
be transformed into variables in the RBR whenever possible). These
annotations cannot be generated when the memory address is not
statically known, though, (for instance, when we have an array access inside a loop
with a variable index). In such cases, we annotate the corresponding
instructions with ``?''.

Finally, when we have Solidity code available, we are able to
retrieve the name of the functions invoked from the hash codes (see
e.g. Block 152 in which we have annotated in the second bytecode
\textsf{kingBlock}, the name of the function to be invoked).  This
allows us to statically know the continuation block.






{\setlength\belowcaptionskip{-10pt}{
\begin{figure}[t]
  \begin{minipage}{0.4\textwidth}
  
\begin{tabular}{|l|}
\hline
\hspace{3pt}
\begin{lstlisting}[name=sol,basicstyle=\scriptsize,keywordstyle=\bfseries\scriptsize]
contract BlockKing {
  $\cdots$
  uint public warriorBlock;
  uint public kingBlock;
  $\cdots$
  function kingBlock(){
    uint var = kingBlock;
    $\cdots$
  }

  function process_payment() {
    uint singleDigit = warriorBlock;
    $\cdots$
    while (singleDigit > 10) {
      singleDigit -= 10;
    }
    $\cdots$
  }
}

\end{lstlisting}\\ 
\hline 
\end{tabular}


  \end{minipage}
  \hspace{0.3cm}
  \begin{minipage}{0.6\textwidth}
  \includegraphics[scale=0.37]{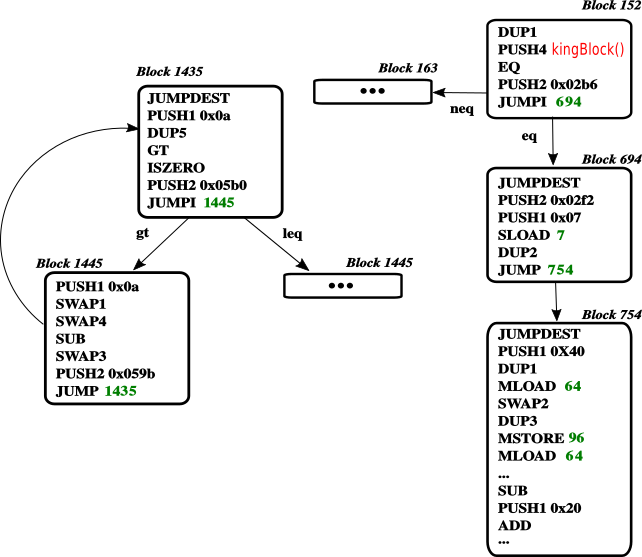}
  \end{minipage} \vspace{-0.3cm}
\caption{Solidity code (left), and EVM code for
  \code{process_payment} within CFG (right).}
\label{fig:cfg}
\end{figure}
}}

\subsection{From the CFG to Guarded Rules}

The translation from EVM into our \emph{rule-based representation} is
done by applying the translation in Def.~\ref{translate} to each block
in a CFG. The identifiers given to the rules --\emph{block\_x} or
\emph{jump\_x}-- use \emph{x}, the PC of the first bytecode in the
block being translated. 
We distinguish among three cases: (1) if the last 
bytecode in the block is an unconditional jump (\textsf{JUMP}), we
generate a single rule, with an invocation to the continuation block,
(2) if it is a conditional jump (\textsf{JUMPI}) we produce two
additional \emph{guarded} rules which represent the continuation when
the condition holds, and when it does not, (3) otherwise, we continue the
execution in block \emph{x+s} (where \emph{s} is the size of the
EVM bytecodes in the block being translated).
As regards the variables, we distinguish the following types:

\begin{enumerate}

\item \emph{Stack variables}: a key ingredient of the translation is that the stack
is flattened into variables, i.e., the part of the stack that the
block is using is represented, when it is translated into a rule, by
the explicit variables $s_0,s_1,\ldots$, where $s_1$ is above $s_0$,
and so on. The initial stack variables are obtained as parameters $s_0,s_1,\ldots,s_n$ and denoted as $\bar{s}_n$.

\item \emph{Local variables}: the content of the local memory in
  numeric addresses appearing in the code, which are accessed through
  MSTORE and MLOAD with the given address, are modelled with variables
  $l_0,l_1,\ldots,l_r$, denoted as $\bar{l}_r$, and are passed as
  parameters. For the translation, we assume we are given a map
  \texttt{lmap} which associates a different local variable to every
  numeric address memory used in the code. When the address is not
  numeric, we represent it using a
  fresh variable local to the rule to indicate that we do not have
  information on this memory location.

\item \emph{Contract fields}: we model fields with variables
  $g_0,\ldots,g_k$, denoted as $\bar{g}_k$, which are passed as
  parameters. Since these fields are accessed using SSTORE and SLOAD
  using the number of the field, we associate $g_i$ to the $i$th
  field. As for the local memory, if the number of the field is not
  numeric because it is unknown (annotated as ``?''), we use a fresh local variable to represent it.

\item \emph{Blockchain data}: we model this data with variables
  $\overline{bc},$ which are either indexed with $md_0,\ldots,md_q$ when
  they represent the message data, or with  corresponding names, if
  they are precise information of the call, like the gas, which is
  accessed with the opcode GAS, or about the blockchain, like the
  current block number, which is accessed with the opcode NUMBER. All
  this data is accessed through dedicated opcodes, which may consume
  some offsets of the stack and normally place the result on top of
  the stack (although some of them, like CALLDATACOPY, can store
  information in the local memory).
 \end{enumerate}
 The translation uses an auxiliary function $\translate$ to
 translate each bytecode into  corresponding high-level instructions
 (and updates the size of the stack $m$)
 and $\translate_G$ to translate the guard of a conditional jump. The
 grammar of the resulting RBR language  into which the EVM is
 translated is given in Fig.~\ref{fig:grammar}. We optionally can keep
 in the RBR the original bytecode instructions from which the
 higher-level ones are obtained by simply wrapping them within a
 \textsf{nop} functor (e.g., \textsf{nop(DUPN)}). This is relevant for
 a gas analyzer to assign the precise gas consumption to the
 higher-level instruction in which the bytecode was transformed.

\begin{definition}\label{translate}
Given a block B with instructions $b_1,\ldots,b_i$ in a CFG starting
at PC $x,$ and local variables map \texttt{lmap}, the generated rules
are:
{\small
\[
\begin{array}{|lll|} \hline
\multicolumn{3}{|c|}{$if $b_i \equiv \textsf{JUMP}~ p }\\
block\_x(\bar{s}_n,\bar{g}_k,\bar{l}_r,\bar{bc_q})  & \Rightarrow &
\translate(b_1,\ldots,b_{i-1}),call(block\_p(\bar{s}_{m-1},\bar{g}_k,\bar{l}_r,\bar{bc_q}))
\\ \hline
\multicolumn{3}{|c|}{$if $b_i \equiv \textsf{JUMPI}~ p }\\
block\_x(\bar{s}_n,\bar{g}_k,\bar{l}_r,\bar{bc_q}) & \Rightarrow&
\translate(b_1,\ldots,b_{c-1}),call(jump\_x(\bar{s}_m,\bar{g}_k,\bar{l}_r,\bar{bc_q}))\\
jump\_x(\bar{s}_n,\bar{g}_k,\bar{l}_r,\bar{bc_q}) &\Rightarrow&
 \translate_G(b_c,\ldots,b_{i-2}) |call(block\_p(\bar{s}_m,\bar{g}_k,\bar{l}_r,\bar{bc_q}))\\
jump\_x(\bar{s}_n,\bar{g}_k,\bar{l}_r,\bar{bc_q}) &\Rightarrow&
\neg\translate_G(b_c,\ldots,b_{i-2}) |call(block\_(x+s)(\bar{s}_m,\bar{g}_k,\bar{l}_r,\bar{bc_q})) \\ \hline
\multicolumn{3}{|c|}{$if $b_i \not\equiv \textsf{JUMP}$ and $b_i \not\equiv \textsf{JUMPI}} \\
block\_x(\bar{s}_n,\bar{g}_k,\bar{l}_r,\bar{bc_q})  & \Rightarrow &
\translate(b_1,\ldots,b_i),call(block\_(x+i)(\bar{s}_m,\bar{g}_k,\bar{l}_r,\bar{bc_q}))
\\ \hline
\end{array}
\]
} where functions $\translate$ and $\translate_G$ for some
representative bytecodes are: {\small
\[
\begin{array}{|lll|} \hline
\translate(\textsf{JUMPDEST}) & = & \{\};~m := m \\ 
\translate(\textsf{PUSHN}~v) & = & \{s_{m+1}= v\};~m := m+1 \\ 
\translate(\textsf{DUPN}) & = & \{s_{m+1}= s_{m+1-N}\};~m := m+1 \\
\translate(\textsf{SWAPN}) & = & \{s_{m+1}= s_m, s_m = s_{m-N},s_{m-N}=s_{m+1}\};~m := m \\
\translate(\textsf{ADD}|\textsf{SUB}|\textsf{MUL}|\textsf{DIV}) & = & \{s_{m-1}= s_m +|-|*|/ s_{m-1}\};~m := m-1 \\
\translate(\textsf{SLOAD}|\textsf{MLOAD}~v) & = & \{s_m= g_v|l_{lmap(v)}\};~m := m~~~~~~~~~~~~~~~~~~\mbox{ if } v \mbox{ is numeric}\\ 
& = & \{gl|ll = s_m, s_m= \textsf{fresh}()\};~m := m~~~~~~~~~~~~~~~\mbox{ otherwise}\\
\translate(\textsf{SSTORE}|\textsf{MSTORE}~v) & = & \{g_v|l_{lmap(v)}= s_{m-1}\};~m := m-2~~~~~~~~~~\mbox{ if } v \mbox{ is numeric}\\
& = & \{gs_1|ls_1= s_{m-1},gs_2|ls_2= s_m\};~m := m-2~~~~\mbox{otherwise}\\
\ldots & &
\\ \hline
\translate_G(\textsf{GT,ISZERO})| \translate_G(\textsf{GT}) & = & leq(s_m,s_{m-1})|gt(s_m,s_{m-1});~m := m-2 \\ 
\translate_G(\textsf{EQ,ISZERO})|\translate_G(\textsf{EQ}) & = & neq(s_m,s_{m-1})|eq(s_m,s_{m-1});~m := m-2 \\
\ldots & &
\\ \hline
\end{array}
\]
}
\begin{itemize}
\item 
  $c$ is the index of the instruction, where the guard of the conditional jump starts. Note that the condition ends at the index $i-2$ and there is always a $\textsf{PUSH}$ at $i-1$. Since the pushed address (that we already have in $p$) and the result of the condition are consumed by the $\textsf{JUMPI}$, we do not store them in stack variables.
\item 
  $m$ represents the size of the stack for the block. Initially we have $m:=n$.
\item variables $gs_1$, $gs_2$ and $gl$, and $ls_1$, $ls_2$ and $ll$, are local to each rule and are used to represent the use of $\textsf{SLOAD}$ and $\textsf{SSTORE}$, and $\textsf{MLOAD}$ and $\textsf{MSTORE}$, when the given address is not a concrete number. For $\textsf{SLOAD}$ and $\textsf{MLOAD}$ we also use $\textsf{fresh}()$, to denote a generator of fresh variables to safely represent the unknown value of the loaded address. 
\end{itemize}
\end{definition}

{\setlength\belowcaptionskip{-10pt}{
\begin{figure}[t]
{\small{
\[
\begin{array}{|lll|} \hline
  RBR & {\rightarrow} &  (B\mid J)~~RBR \mid \epsilon\\

  B & {\rightarrow} & block\_{id}~(\inot{n},\got{k},\lot{r},\blot) \Rightarrow Instr~~(Call \mid \epsilon)\\[1pt]
  
  J & {\rightarrow} &  jump\_{id}~(\inot{n},\got{k},\lot{r},\blot) \Rightarrow InstrJ\\[1pt]

  Instr & {\rightarrow} & S~~Instr \mid \epsilon\\[1pt]


  S &{\rightarrow}& s = Exp \\[1pt]
  
  Exp &{\rightarrow} & num \mid x \mid x+y \mid x-y \mid x*y \mid x/y
                       \mid x\%y \mid x^y
  \\[1pt]
      &&\mid and(x,y) \mid or(x,y) \mid xor(x,y) \mid not(x) \\[1pt]
  
  Call &{\rightarrow}& call(block\_{id}(\inot{n},\got{k},\lot{r},\blot)) \mid
                    call(jump\_{id}(\inot{n},\got{k},\lot{r},\blot))\\[1pt]
  InstrJ & {\rightarrow} & Guard~"|"~call(block\_{id}(\inot{n},\got{k},\lot{r},\blot))\\[1pt]
  Guard & {\rightarrow} & eq(x,y) \mid neq(x,y) \mid lt(x,y) \mid
                          leq(x,y) \mid gt(x,y) \mid geq(x,y)\\[1pt] \hline

\end{array}
\] 
}}
\caption{Grammar of the RBR into which the EVM is translated} \label{fig:grammar}
\end{figure}
}}

\secbeg\secbeg\secbeg
\begin{example}
As an example, an excerpt of the 
RBR obtained by translating the three blocks on the right-hand side of
Fig.~\ref{fig:cfg} is as follows (selected instructions keep using
\textsf{nop} annotations the bytecode from which they have been obtained):

\setlength{\tabcolsep}{0.5pt}
\noindent
\begin{tabular}[t]{|l|l|l|}
\hline
\hspace{4pt}
\begin{lstlisting}[name=deco,escapeinside={/@}{@/}]
$block152(\svar{0},\got{11},\lot{8},\blot)\Rightarrow$
  $\svar{1} = \svar{0}~_{\textsf{nop(DUP1)}}$
  $\svar{2} = 6584849474~_{\textsf{nop(PUSH4)}}$
  $call(jump152(\sot{2},\got{11},\lot{8},\blot)$/@\vspace{2pt}@/ 
  $_{\textsf{nop(EQ)}}~_{\textsf{nop(PUSH2)}}~_{\textsf{nop(JUMPI)}}$
$jump152(\sot{2},\got{11},\lot{8},\blot)\Rightarrow$
  $eq(\svar{2},\svar{1})$
  $call(block694(\svar{0},\got{11},\lot{8},\blot)$/@\vspace{2pt}@/
$jump152(\sot{2},\got{11},\lot{8},\blot)\Rightarrow$
  $neq(\svar{2},\svar{1})$
  $call(block163(\svar{0},\got{11},\lot{8},\blot)$
\end{lstlisting}
~~&
\hspace{4pt}
\begin{lstlisting}[name=deco,escapeinside={/@}{@/}]%,lineskip=1pt%,framextopmargin=2pt]
  
$block694(\svar{0},\got{11},\lot{8},\blot)\Rightarrow$
  $\svar{1} = 754~_{\textsf{nop(PUSH2)}}$
  $\svar{2} = 7~_{\textsf{nop(PUSH1)}}$
  $\svar{2} = g_7~_{\textsf{nop(SLOAD)}}$
  $\svar{3}=  \svar{1}~_{\textsf{nop(DUP2)}}$
  $call(block754(\sot{2},\got{11},\lot{8},\blot)$/@\vspace{2pt}@/
  $_{\textsf{nop(JUMP)}}$
$block754(\sot{2},\got{11},\lot{8},\blot)\Rightarrow$
  $\svar{3} = 64~_{\textsf{nop(PUSH1)}}$
  $\svar{4} = \svar{3}~_{\textsf{nop(DUP1)}}$
  $\svar{4} = \lvar{0}~_{\textsf{nop(MLOAD)}}$
\end{lstlisting}
~~&
\hspace{4pt}
\begin{lstlisting}[name=deco]%,framextopmargin=2pt]

$\svar{5} = \svar{4}$
$\svar{4} = \svar{2}$
$\svar{2} = \svar{5}~_{\textsf{nop(SWAP2)}}$
$\svar{5} = \svar{2}~_{\textsf{nop(DUP3)}}$
$\lvar{1} = \svar{4}~_{\textsf{nop(MSTORE)}}$
$\svar{3} = \lvar{0}~_{\textsf{nop(MLOAD)}}$
$\cdots$
$\svar{3} = \svar{4}-\svar{3}~_{\textsf{nop(SUB)}}$
$\svar{4}= 32~_{\textsf{nop(PUSH1)}}$
$\svar{3} = \svar{4}+\svar{3}~_{\textsf{nop(ADD)}}$
$\cdots$
\end{lstlisting}\\
\hline
\end{tabular}

\end{example}








\section{Case Study: Bounding Loops in  EVM using \tname{SACO}}
\label{sec:analysis}

To illustrate the applicability of our framework, we have analyzed
quantitative properties of EVM code by translating it into our
intermediate representation and analyzing it with the high-level static analyzer
SACO \cite{AlbertAFGGMPR14}. SACO 
is able to infer, among other properties,  \emph{upper bounds} on
the number of iterations of loops. Note that this is the first crucial step to infer the gas
consumption of smart contracts, a property of much interest~\cite{Eth-gas}.
%
The internal representation of SACO (described
in\cite{AlbertACGGPR15}) matches the grammar in Fig.~\ref{fig:grammar}
after minor syntactic translations (that we have solved implementing a
simple translator that is available in github, named \texttt{saco.py}). As
SACO does not have bit-operations (namely \texttt{and}, \texttt{or},
\texttt{xor}, and \texttt{not}), our translator replaces such
operations by fresh variables so that the analyzer forgets the
information on bit variables.  After this, for our running example, we
prove termination of the 6 loops that it contains and produce a linear
bound for those loops. We have included in our github other smart
contracts together with the loop bounds inferred by SACO for
them. Other high-level analyzers that work on intermediate forms like
Integer transition systems or Horn clauses  (e.g.,
AproVe, T2, VeryMax, CoFloCo) could be easily adapted as well to
work on our RBR translated programs. 

%




\section{Related Approaches and Tools}
\label{sec:related}

In the past two years, several approaches tackled the challenge of
fully formal reasoning about Ethereum contracts implemented directly
in EVM bytecode by modeling its rigorous semantics in
state-of-the-art proof
assistants~\cite{Hirai:WTSC17,Grishchenko-al:POST18}.
While those mechanisations enabled formal machine-assisted proofs of
various safety and security properties of EVM
contracts~\cite{Grishchenko-al:POST18}, none of them provided means
for fully \emph{automated} analysis of EVM bytecode.

Concurrently, several other approaches for ensuring correctness and
security of Ethereum contracts took a more agressive approach,
implementing automated toolchains for detecting bugs by symbolically
executing EVM
bytecode~\cite{Luu-al:CCS16,Nikolic-al:Maian}.
%
%
However, low-level EVM representation poses difficulties in applying
those tools immediately for analysis of more high-level properties.
For instance, representation of EVM in \oyente, a popular tool for
analysis of Ethereum smart contracts~\cite{Oyente} is too low-level to
implement analyses of high-level properties, \eg, loop complexity or
commutativity conditions.
\tname{Porosity}~\cite{porosity} is a decompiler from (a subset of)
EVM directly into \tname{Solidity}. 
Since \tname{Solidity} does not
have a formal semantics, it is not well-suited for performing program
analyses.






\bibliographystyle{abbrv}
\bibliography{biblio}

\end{document}